\PassOptionsToPackage{unicode}{hyperref}
\PassOptionsToPackage{hyphens}{url}
\PassOptionsToPackage{dvipsnames,svgnames,x11names}{xcolor}
\documentclass[
  12pt]{article}

\usepackage{amsmath,amssymb}
\usepackage{iftex}
\ifPDFTeX
  \usepackage[T1]{fontenc}
  \usepackage[utf8]{inputenc}
  \usepackage{textcomp} 
\else 
  \usepackage{unicode-math}
  \defaultfontfeatures{Scale=MatchLowercase}
  \defaultfontfeatures[\rmfamily]{Ligatures=TeX,Scale=1}
\fi
\usepackage{lmodern}
\ifPDFTeX\else  
\fi
\IfFileExists{upquote.sty}{\usepackage{upquote}}{}
\IfFileExists{microtype.sty}{
  \usepackage[]{microtype}
  \UseMicrotypeSet[protrusion]{basicmath} 
}{}
\makeatletter
\@ifundefined{KOMAClassName}{
  \IfFileExists{parskip.sty}{%
    \usepackage{parskip}
  }{
    \setlength{\parindent}{0pt}
    \setlength{\parskip}{6pt plus 2pt minus 1pt}}
}{
  \KOMAoptions{parskip=half}}
\makeatother
\usepackage{xcolor}
\makeatletter
\ifx\paragraph\undefined\else
  \let\oldparagraph\paragraph
  \renewcommand{\paragraph}{
    \@ifstar
      \xxxParagraphStar
      \xxxParagraphNoStar
  }
  \newcommand{\xxxParagraphStar}[1]{\oldparagraph*{#1}\mbox{}}
  \newcommand{\xxxParagraphNoStar}[1]{\oldparagraph{#1}\mbox{}}
\fi
\ifx\subparagraph\undefined\else
  \let\oldsubparagraph\subparagraph
  \renewcommand{\subparagraph}{
    \@ifstar
      \xxxSubParagraphStar
      \xxxSubParagraphNoStar
  }
  \newcommand{\xxxSubParagraphStar}[1]{\oldsubparagraph*{#1}\mbox{}}
  \newcommand{\xxxSubParagraphNoStar}[1]{\oldsubparagraph{#1}\mbox{}}
\fi
\makeatother

\usepackage{longtable,booktabs,array}
\usepackage{calc} 
\usepackage{etoolbox}
\makeatletter
\patchcmd\longtable{\par}{\if@noskipsec\mbox{}\fi\par}{}{}
\makeatother
\IfFileExists{footnotehyper.sty}{\usepackage{footnotehyper}}{\usepackage{footnote}}
\makesavenoteenv{longtable}
\usepackage{graphicx}
\makeatletter
\def\maxwidth{\ifdim\Gin@nat@width>\linewidth\linewidth\else\Gin@nat@width\fi}
\def\maxheight{\ifdim\Gin@nat@height>\textheight\textheight\else\Gin@nat@height\fi}
\makeatother
\setkeys{Gin}{width=\maxwidth,height=\maxheight,keepaspectratio}
\makeatletter
\def\fps@figure{htbp}
\makeatother

\makeatletter
\@ifpackageloaded{caption}{}{\usepackage{caption}}
\AtBeginDocument{%
\ifdefined\contentsname
  \renewcommand*\contentsname{Table of contents}
\else
  \newcommand\contentsname{Table of contents}
\fi
\ifdefined\listfigurename
  \renewcommand*\listfigurename{List of Figures}
\else
  \newcommand\listfigurename{List of Figures}
\fi
\ifdefined\listtablename
  \renewcommand*\listtablename{List of Tables}
\else
  \newcommand\listtablename{List of Tables}
\fi
\ifdefined\figurename
  \renewcommand*\figurename{Figure}
\else
  \newcommand\figurename{Figure}
\fi
\ifdefined\tablename
  \renewcommand*\tablename{Table}
\else
  \newcommand\tablename{Table}
\fi
}
\@ifpackageloaded{float}{}{\usepackage{float}}
\floatstyle{ruled}
\@ifundefined{c@chapter}{\newfloat{codelisting}{h}{lop}}{\newfloat{codelisting}{h}{lop}[chapter]}
\floatname{codelisting}{Listing}

\makeatother
\makeatletter
\@ifpackageloaded{caption}{}{\usepackage{caption}}
\@ifpackageloaded{subcaption}{}{\usepackage{subcaption}}
\makeatother
\newtheorem{theorem}{Theorem}
\usepackage[linesnumbered,boxed,ruled,commentsnumbered]{algorithm2e}
\ifLuaTeX
  \usepackage{selnolig}  
\fi
\usepackage[]{natbib}
\usepackage{bookmark}
\newcommand{\equalcontrib}{\thanks{Kuangnan Fang and Ruixuan Qin are co-first authors.}}
\IfFileExists{xurl.sty}{\usepackage{xurl}}{} 
\hypersetup{
  pdftitle={Title},
  pdfauthor={Author 1; Author 2},
  pdfkeywords={3 to 6 keywords, that do not appear in the title},
  colorlinks=true,
  linkcolor={blue},
  filecolor={Maroon},
  citecolor={Blue},
  urlcolor={Blue},
  pdfcreator={LaTeX via pandoc}}

\newcommand{\anon}{1}

\begin{document}

\def\spacingset#1{\renewcommand{\baselinestretch}%
{#1}\small\normalsize} \spacingset{1}


\if1\anon
{
  \title{\bf Transfer learning under latent space model}
 \author{%
Kuangnan Fang\\
 School of Economics,
 Xiamen University\\
  Ruixuan Qin\equalcontrib\\
  School of Economics,
 Xiamen University\\
  and\\
  Xinyan Fan\thanks{Xinyan Fan is the corresponding author. Email address: 20198102@ruc.edu.cn} \\
  Center of Applied Statistics and School of Statistics, \\
  Renmin University of China \\
}
  \maketitle
} \fi

\if0\anon
{
  \bigskip
  \bigskip
  \bigskip
  \begin{center}
    {\LARGE\bf Title}
\end{center}
  \medskip
} \fi

\bigskip
\begin{abstract}
 Latent space model plays a crucial role in network analysis, and accurate estimation of latent variables is essential for downstream tasks such as link prediction. 
   However, the large number of parameters to be estimated presents a challenge, especially when the latent space dimension is not exceptionally small. 
    In this paper, we propose a transfer learning method that leverages information from networks with latent variables similar to those in the target network, thereby improving the estimation accuracy for the target.
    Given transferable source networks, we introduce a two-stage transfer learning algorithm that accommodates differences in node numbers between source and target networks.
     In each stage, we derive sufficient identification conditions and design tailored projected gradient descent algorithms for estimation. Theoretical properties of the resulting estimators are established. When the transferable networks are unknown, a detection algorithm is introduced to identify suitable source networks.
     Simulation studies and analyses of two real datasets demonstrate the effectiveness of the proposed methods.
\end{abstract}

\noindent%
\vfill

\newpage
\spacingset{1.8} 

\section{Introduction}

Network data, which describes relationships between different entities, has garnered significant attention and found widespread applications in various fields, including sociology, biology, and economics \citep{maharjan2022resilient,liu2020computational,kurbucz2023analysis,mostafa2024social}.
Among the various approaches proposed to analyze network data, network embedding methods have proven particularly essential \citep[see, e.g.,][]{lan2020node,yan2023reconciling,iyer2024non}. One representative approach is the latent space model and its extensions. They assume that each node can be embedded in a low-dimensional Euclidean space, where nodes with closer latent positions are more likely to be connected \citep[see, e.g.,][]{hoff2002latent,ma2020universal,zhang2020flexible,zhang2022joint}.
Accurate estimation of latent variables is crucial in the latent space model framework for downstream tasks such as link prediction and community detection \citep{zhang2022joint,fan2024generalized}.
However, the latent space model involves many parameters, especially when the latent space is not very low-dimensional. This makes it challenging to accurately estimate a single network.

With advances in technology, data have become richer and more accessible, allowing us to observe various types of relationships among individuals.
For example, relationships in social networks can include friendships and co-worker connections \citep{zhang2020flexible}, and trade networks between countries can be observed for different products \citep{fan2022alma}. 
Previous research has investigated numerous methods based on the latent space model to leverage multiple networks \citep[see, e.g.,][]{gollini2016joint,zhang2020flexible, macdonald2022latent}. However, their primary objective is the joint estimation of latent variables across all networks based on inter-layer similarities. As a result, these methods are not designed to enhance the accuracy of a specific target network. 
Furthermore, they can only be applied when the node sets across network layers are identical and do not account for variations in the number of nodes between layers.

Transfer learning \citep{torrey2010transfer} is a powerful method to transfer knowledge from related but distinct source datasets to improve task performance in the target dataset.
It has been widely explored within statistical frameworks, such as linear regression models \citep{li2022transfer, he2024transfusion}, generalized linear regression models \citep{tian2023transfer, li2024estimation}, and Gaussian graphical models \citep{li2023transfer, ren2024transfer}.
Transfer learning of network data has been extensively investigated \citep[see, e.g.,][]{he2009graph,tang2016transfer,dong2019transfer,zou2021transfer,wu2023non}, with most studies situated in the field of computer science and lacking theoretical results.
Under the latent space model framework, \cite{jalan2024transfer} proposed a method to identify similar nodes for each node in the target network using the $h$-quantiled smallest distances, which are calculated based on the edges in the source network.
Despite the method's well-behaved nature, it may suffer from two potential drawbacks. First, it considers only one source network for knowledge transfer, whereas in real-world scenarios, multiple networks representing various relationships are often available. Second, the estimation of latent variables in the target network is neglected, as the method primarily focuses on estimating the connecting probability matrix. This omission may limit the interpretability and further analysis of the method.

In this paper, we propose transfer learning methods for latent space model aimed at aggregating knowledge from multiple related but distinct source networks and transferring this knowledge to the target network to enhance parameter estimation and further tasks. When the set of effective source networks is known, we propose a two-stage transfer learning algorithm, denoted as TLK (\textbf{T}ransfer \textbf{l}earning with \textbf{k}nown transferable set). The first stage aggregates information from multiple source networks, and the second stage transfers this information to
estimate the latent variables in
the target network through a debiasing procedure.
To implement the method, we face three challenges: First, the source networks may vary in size, complicating the aggregation of useful information for transfer. Second, unlike traditional regression problems, the process focuses on debiasing the matrix parameters, with more complex matrix norms, such as the nuclear norm, serving as penalty functions. This, in turn, increases computational complexity. Third, due to the presence of parameters in various forms, such as vectors and matrices, studying their theoretical properties is far more challenging than in traditional transfer regression.
To tackle the first challenge, we assume that all the source networks contain the nodes in the target network. 
Carefully designed conditions for identifiability are provided. 
To address the second challenge, we propose an algorithm combining projected gradient descent and proximal gradient descent.
Furthermore, inequalities related to matrix inner products and norms are utilized to investigate the upper bounds of the estimation error. When the set of effective source networks is not known,
we develop a transfer learning algorithm, TLD (\textbf{T}ransfer \textbf{l}earning with \textbf{d}etected transferable set), which first selects the transferable networks from the source networks and then performs knowledge transfer.

Our contributions can be summarized in the following five aspects: First, we propose a novel transfer learning framework to leverage information from multiple source networks.
Second, sufficient conditions for identifiability are provided to accommodate source networks with varying sample sizes.
Third,  feasible algorithms are designed for the identification of transferable networks and the two-stage network transfer process.
Fourth, the estimation properties of the latent position vectors of nodes in the network are rigorously established.
Finally, extensive simulation studies and two real-data applications—on the FAO trade networks and the POLECAT political event networks—are conducted to demonstrate the effectiveness of the proposed methods.

\section{Models and notations}

Consider a target network $\mathcal{G}^t=(\mathcal{V}^t,\mathcal{E}^t)$, where $\mathcal{V}^t=\{v_1^t,\ldots,v_n^t\}$ denotes the node set and $\mathcal{E}^t$ denotes the edge set.
Assume that $\mathcal{G}^t$ is undirected with no self- loops. Then, it can be represented by a symmetric binary adjacency matrix $A^t=(A_{ij}^t)\in \{0,1\}^{n\times n}$, where $A^t_{ij}=A^t_{ji}=1$ if node $i$ and node $j$ are connected, and $A^t_{ij}=A^t_{ji}=0$ otherwise.
Following \cite{hoff2002latent}, we consider a latent space model, which describes each node with a low-dimensional latent vector $Z_i^t\in\mathbb{R}^k$ for $i=1,\ldots,n$.
Given the latent variables, the connecting probability between $v_i^t$ and $v_j^t$ is assumed to be independently Bernoulli distributed with
\begin{equation}  \label{s2-1}
A_{ij}^t\sim \text{Bernoulli}(P_{ij}^t),\quad
        \Theta_{ij}^t=\text{logit}(P_{ij}^t)=\alpha_i^t+\alpha_j^t+Z_i^{t\top }Z^t_j,
\end{equation}
where $\text{logit}(x)=\text{log}\{x/(1-x)\}$, $\alpha_i^t$ corresponds to the degree heterogeneity parameter of node $i$. 
In general,
a larger value of $\alpha_i^t$ indicates a stronger tendency of node $i$ to form connections with other nodes. In addition, for any $i = 1, \ldots, n$, $Z_i^{t}$ denotes the latent representation of node $i$ in the target network. The similarity between nodes $i$ and $j$ is quantified by the inner product $Z_i^{t\top} Z_j^{t}$.
Equation (\ref{s2-1})
can be rewritten into a matrix form 
$
	\Theta^t=\alpha^t 1_n^\top+1_n\alpha^{t\top}+Z^tZ^{t\top},
$
with $\Theta^t=(\Theta_{ij}^t)\in\mathbb{R}^{n\times n}$, $\alpha^t=(\alpha_1^t,\ldots,\alpha_n^t)^\top\in \mathbb{R}^n$, $Z^t=(Z_1^t,\ldots,Z_n^t)^\top\in\mathbb{R}^{n\times k}$, and $1_n=(1,\ldots,1)^\top\in\mathbb{R}^n$.

Besides the target network, consider $L$ source networks, denoted by $\mathcal{G}^{s_1},\ldots,\mathcal{G}^{s_L}$. For each $s_l\in \{s_1,\ldots,s_L\}$, the network $\mathcal{G}^{s_l}$ consists of $N_l$ nodes $\{v_1^l,\dots,v_{N_l}^l\}$. 
In this paper, we assume that the node set of the target network is a subset of the node set of each source network, i.e., $\{v_1^t,\dots,v_n^t\}\subseteq \{v_1^l,\dots,v_{N_l}^l\}$ for $l=1,\dots,L$. The number of nodes $N_l$ is allowed to vary across networks.
Let $A^{s_1}, \ldots, A^{s_L}$ denote the adjacency matrices corresponding to the $L$ source networks, where each $A^{s_l} = (A_{ij}^{s_l}) \in \{0,1\}^{N_l \times N_l}$ for $i, j = 1, \ldots, N_l$ and $l = 1, \ldots, L$.
For the $l$-th source network, 
\begin{equation}\label{equ S}
  A_{ij}^{s_l}\sim \text{Bernoulli}(P_{ij}^{s_l}),  \quad 
	\Theta_{ij}^{s_l}=\text{logit}(P_{ij}^{s_l})=\alpha_i^{s_l}+\alpha_j^{s_l}+Z_i^{s_l\top }Z^{s_l}_j,
\end{equation}
where $Z_i^{s_l}\in\mathbb{R}^{k}$ for $i=1,\ldots,N_l$ representing the latent variable of node $i$ in the $l$-th source network, and $\alpha_i^{s_l}$ is the degree heterogeneity parameter for node $i$.
The matrix form is 
$
	\Theta^{s_l}=\alpha^{s_l} 1_{N_l}^\top+1_{N_l}\alpha^{s_l\top}+Z^{s_l}Z^{s_l\top},
$
 with $\Theta^{s_l}=(\Theta_{ij}^{s_l})\in \mathbb{R}^{N_l\times N_l}$, $\alpha^{s_l}=(\alpha_1^{s_l},\ldots,\alpha_{N_l}^{s_l})^\top\in\mathbb{R}^{N_l}$, $Z^{s_l}=(Z_1^{s_l},\ldots,Z_{N_l}^{s_l})^\top\in\mathbb{R}^{N_l\times k}$.
Without loss of generality, we assume that the  first $n$ nodes in each source network correspond to the nodes in target network, maintaining the same order. Thus, we partition the latent variable $Z^{s_l}$ into two parts: $Z^{s_l}=(U_{0l}^\top,U_l^\top)^\top$, where $U_{0l} \in \mathbb{R}^{n \times k}$ represents the matrix of latent variables for the first $n$ nodes (i.e., the nodes in the target network), while $U_l \in \mathbb{R}^{n_l \times k}$ corresponds to the matrix of latent variables for the remaining nodes not included in the target network. The total number of nodes satisfies $n+n_l=N_l$.

To assess the similarity between source networks and the target network, 
we define the informative set given the transferring level $\delta\in\mathbb{R}^+$ as 
\begin{equation}\label{equ A}
    \mathcal{A}\equiv\mathcal{A}_\delta=\left\{l\in\{0,\ldots,L\}:\lVert U_{0l}-Z^t\rVert_*\leq \delta\right\}.
\end{equation}
The size of $\mathcal{A}$ is determined  by the value of $\delta$. A large value of $\delta$ allows more source networks to be considered in the transfer learning process, while a smaller $\delta$ selects source networks with latent variables more similar to those of the target network. 
In this paper, we chose the nuclear norm to measure the distance of $U_{0l}$ and $Z^t$. The nuclear norm of a matrix, defined as the sum of its singular values, captures information across multiple dimensions of the matrix. Other matrix norms, such as the spectral norm and Frobenius norm, can also be considered in the further studies.

\section{Transfer learning for networks}
In this section,
we first assume transferable set $\mathcal{A}$ is known, and
 propose a two-stage transfer learning algorithm. 
Additionally, we investigate the theoretical properties of the estimator. At last, we propose an algorithm to detect the transferable set.

\subsection{Two-stage transfer learning algorithm with $\mathcal A$ known}

The transfer learning algorithm for the latent space model is inspired by the methods of \cite{li2022transfer} and \cite{tian2023transfer}.
In the first stage, we compute an initial estimator of the  matrix of latent variables for the target network using information from all transferable source networks. In the second stage, we apply a debiasing procedure through penalization to correct the bias introduced by the differences in latent matrices between the source and target networks.

\textbf{Transferring stage.}
 We denote the aggregated matrix of latent variables for the transferable source networks as $U_0$, which encapsulates the knowledge provided by the source data for the target network.
Note that $U_0\in\mathbb{R}^{n\times k}$ is also an approximate  estimator of $Z^{t}$.
To derive an estimator of $U_0$, we consider
 the minimization of the following negative log-likelihood function,
\begin{equation}  \label{2-2}
	\mathcal L(U_0,\{\alpha^{s_l},U_l\}_{l=1}^{|\mathcal A|})=-\sum_{l\in\mathcal{A}}\sum_{i=1}^{N_l}\sum_{j=1}^{N_l}\left\{A_{ij}^{s_l}\mathring\Theta_{ij}^{s_l}+\text{log}\{1-\sigma(\mathring\Theta_{ij}^{s_l})\}\right\},
\end{equation}
where $\sigma(x)=1/(1+\text{exp}(-x))$, and $ \mathring\Theta^{s_l}=\alpha^{s_l} 1_{N_l}^\top+1_{N_l}\alpha^{s_l\top}+\mathring Z^{s_l}\mathring Z^{s_l\top}$, $\mathring Z^{s_l}=(U_0^\top,U_l^\top)^\top$. 
The diagonal elements of each network are included in the objective function,  following the result in \cite{ma2020universal}, which showed that their impact is negligible in both theory and practice.

It is evident that the minimization of (\ref{2-2}) is not unique due to the indeterminacies resulting from the addition of $\alpha^{s_l}$ and $\mathring Z^{s_l}$ for each $l\in\mathcal{A}$.  Additional constraints on the parameters are necessary. In Proposition 1, we provide the sufficient  conditions for identifiability, and the proof is provided in  the supplementary material.
Denote $J_1=I_n-1/n1_n1_n^\top$, and
$J_2^l=I_{n_l}-1/n_l1_{n_l}1_{n_l}^\top$.

\textbf{Proposition 1}:  
 {\it Assume that 
 $J_1U_0=U_0$, $J_1U_0'=U_0'$, 
 $U_0$ and $U_0'$ are of full column rank,  and
 $J_2^lU_l=U_l$, $J_2^lU_l'=U_l'$ for $l\in\mathcal{A}$, $\lvert\mathring\Theta_{ij}^{s_l}\rvert<\infty$ and $\lvert\mathring\Theta_{ij}^{s_l'}\rvert<\infty$.
 {Then  if
 \begin{equation}\label{3-1}
    -\sum_{l\in\mathcal{A}}\sum_{i=1}^{N_l}\sum_{j=1}^{N_l}\left\{A_{ij}^{s_l}\mathring\Theta_{ij}^{s_l}+\text{log}\{1-\sigma(\mathring\Theta_{ij}^{s_l})\}\right\}=-\sum_{l\in\mathcal{A}}\sum_{i=1}^{N_l}\sum_{j=1}^{N_l}\left\{A_{ij}^{s_l}\mathring\Theta_{ij}^{s_l'}+\text{log}\{1-\sigma(\mathring\Theta_{ij}^{s_l'})\}\right\}
 \end{equation}
 reaches the minimum of (\ref{2-2}), then there exists an orthogonal matrices $O\in \mathbb{R}^{k\times k}$ 
 such that 
 \begin{equation*}\label{s2}
     \alpha^{s_l}=\alpha^{s_l\prime}, U_0=U_0'O, U_l=U_l'O,
 \end{equation*}}
 for $l\in\mathcal{A}$, where  $ \mathring\Theta^{s_l'}=\alpha^{s_l'} 1_{N_l}^\top+1_{N_l}\alpha^{s_l'\top}+(U_0'^\top,U_l'^\top)^\top(U_0'^\top,U_l'^\top)$.}

 In Proposition 1, we assume $J_1U_0=U_0$ in order to distinguish $\mathring Z^{s_l}$ and $(\alpha_1^{s_l},\ldots,\alpha_{n}^{s_l})^\top$ for $l\in\mathcal{A}$. Similarly, $J_2^lU_l=U_l$ is used to separate $\mathring Z^{s_l}$ from $(\alpha_{n+1}^{s_l},\ldots,\alpha_{N_l}^{s_l})^\top$. 
 Based on the sufficient conditions,  $\alpha^{s_l}$ is identifiable for $l\in\mathcal{A}$, and $U_0$, $U_l$ are identifiable up to an orthogonal transformation for all $l\in\mathcal{A}$.
With the identification conditions holding, equation (\ref{2-2}) can be estimated using a projected gradient descent algorithm, detailed in the supplementary Material.

\textbf{Debiasing stage.}
Denote $(\hat U_0,\{\hat \alpha^{s_l},\hat U_l\}_{l=1}^{|\mathcal A|})=\text{argmin}_{J_1U_0=U_0, J_2^lU_l=U_l}\mathcal L(U_0,\{\alpha^{s_l},U_l\}_{l=1}^{|\mathcal A|})$ as the estimators obtained in Transferring stage.
Due to the existence of a distance between  $Z^t$ and $U_{0l}$, the estimated $\hat U_0$ is a biased estimator for the latent variables in the target network. Therefore, a debiasing procedure is necessary.
Incorporating the data from the target network, we consider the following model:
\begin{equation*}
    \Theta^t=\alpha^t 1_n^\top+1_n\alpha^{t\top}+(\hat U_0+\Delta)(\hat U_0+\Delta)^{\top},
\end{equation*}
where $\Delta \in\mathbb{R}^{n\times k}$ captures the bias between $\hat U_0$ and $Z^t$. A penalized objective function is then employed to promote a small nuclear norm of $\Delta$:
\begin{equation}\label{2-4}  
	\underset{J_1\Delta=\Delta}{\text{min}}-\sum_{i=1}^n\sum_{j=1}^n\left\{A_{ij}^t\Theta_{ij}^t+\text{log}\{1-\sigma(\Theta_{ij}^t)\}\right\}+\lambda ||\Delta||_*,
\end{equation}
where $\lambda$ is a tuning parameter, and the identification condition $J_1\Delta=\Delta$ is imposed to ensure the identification of $\hat U_0+\Delta$ up to a rotation, similar to the technique in Proposition 1. 
In this paper, we perform debiasing based on nuclear norm regularization, which aligns with our definition of the transferable network set in (\ref{equ A}). Nuclear norm regularization is commonly used for estimating low-rank matrices. The rank of \( \Delta \) is less than \( k \) implies that the discrepancy between \( U_0 \) and \( Z^t \) lies in a lower-dimensional space.

The estimation of the objective function (\ref{2-4}) is detailed in Algorithm  in the supplementary material. A combined approach using the projected gradient descent algorithm and the proximal gradient descent algorithm is developed for this purpose. The estimated $\hat \Delta$ is then added to $\hat U_0$ to obtain the final estimation of the latent variables in the target network.

We can summarize the two-stage algorithm of TLK in Algorithm \ref{al3}. The proposed algorithms have affordable computational cost.  For example, a simulation with $n=200$, $N=200$, $\lvert\mathcal{A}\rvert=10$, $k=2$ and a candidate set for $\lambda$ of cardinality 11 takes around 4.5 minutes on a computer equipped with Intel(R) Xeon(R) E5-2643 v4 CPUs and 256GB of RAM.
\subsection{Theoretical result}

Define the true parameters of $\Theta^{t}$, $Z^{t}$, and $\alpha^{t}$ as $\Theta^{t*}$, $Z^{t*}$, and $\alpha^{t*}$, respectively.
Let $U_0^{*}$, $\alpha^{s_l*}$'s, $U_l^{*}$'s be the minimizer of ${\rm E}(\mathcal L(U_0,\{\alpha^{s_l},U_l\}_{l=1}^{|\mathcal A|}))$. 
Define  $\Delta^*=Z^{t*}-U_0^*$, and $\delta^{*}=\|\Delta^{*}\|_*$. 
Furthermore, for a given matrix, $U_0$, satisfying $1_n^\top U_0=0$, define the parameter space of target network as 
    \begin{equation*}
	\begin{aligned}
\mathcal{F}_1(U_0)=\{&\left(\alpha^t,\Delta\right):\Theta^t=\alpha^t 1_n^\top+1_n\alpha^{t\top}+( U_0+\Delta)(U_0+\Delta)^{\top},
 J_1\Delta=\Delta, \lVert \Theta^t\rVert_{\text{max}}\leq M, \\&||\Delta||_*\leq \delta^{*}+\xi+C_{d}\sqrt{n}\},
	\end{aligned}
\end{equation*}
where $M$ and $C_d$ are two finite positive constants and $\xi=\text{min}_{O:OO^\top=O^\top O=I_k}\|U_0-U_0^{*}O\|_*$. It is easy to verify that there exist $\alpha^{t}$ and $\Delta$ in $\mathcal{F}_1(U_0)$
such that $\Theta^* = \alpha^t 1_n^\top+1_n\alpha^{t\top}+( U_0+\Delta)(U_0+\Delta)^{\top}$ for $\lVert \Theta^{t*}\rVert_{\text{max}}\leq M$.
Let $(\hat\alpha^{t},\hat\Delta)$ be the optimizer of objective function (\ref{2-4}) under parameter space $\mathcal{F}_1(U_0)$ given $\hat U_0=U_0$. 

\begin{algorithm}[H]
	\caption{Two-stage transfer learning algorithm} \label{al3}
	\SetAlgoNoLine
	\SetKwInOut{Input}{\textbf{Input}}
	\SetKwInOut{Output}{\textbf{Output}}
	
	\Input{transferable set $\mathcal{A}$; adjacency matrices of source networks $A^{s_l} \in \mathbb{R}^{N_l \times N_l}$ for $l\in\mathcal{A}$; adjacency matrix of target network $A^t\in\mathbb{R}^{n\times n}$; latent space dimension $k$; tuning parameter $\lambda$.
	}
\Output{$\hat{\alpha}^{t},\hat{Z}^t.$
	}
	\BlankLine
    
	\textbf{Tranferring stage:} \begin{footnotesize}$\left((\hat\alpha^{s_l})_{l\in\mathcal{A}},\hat U_0,(\hat U_l)_{l\in\mathcal{A}}\right)=\underset{J_1U_0=U_0, J_2^lU_l=U_l}{\text{argmin}}-\sum_{l\in\mathcal{A}}\sum_{i,j=1}^{N_l}\left\{A_{ij}^{s_l}\mathring\Theta_{ij}^{s_l}+ \text{log}\{1-\sigma(\mathring\Theta_{ij}^{s_l})\}\right\}$\end{footnotesize}\;
   \textbf{Debiasing stage:} \begin{footnotesize}$(\hat\alpha^t,\hat\Delta)=\underset{J_1\Delta=\Delta}{\text{argmin}}$ $-\sum_{i,j=1}^n\left\{A_{ij}^t\Theta_{ij}^t+\text{log}\{1-\sigma(\Theta_{ij}^t)\}\right\}+\lambda ||\Delta||_*$\;
$\hat Z^t=\hat U_0+\hat \Delta$ \end{footnotesize}\;
			\textbf{Output} $\hat \alpha^t$ and $\hat Z^t$.

\end{algorithm}

Denote $\zeta_n=\|U_0U_0^{\top}-U_0^{*}U_0^{*\top}\|_F$. For any matrix $B$, let $\kappa_i(B)$ be the $i$-th largest eigenvalue of $B$.
To introduce our theoretical result, following technical conditions are assumed.
\begin{itemize}
\item [(C0)] Assume $\delta^*=O(1)$, and $  k=O(1)$. 
    \item [(C1)]Assume $\xi=O(1)$ and $\zeta_n\leq c_{\zeta}\sqrt{n}$ for some constants $c_{\zeta}>0$. 
    \item [(C2)] There exists two finite positive constants $\nu_1<\nu_2$ such that $\nu_1<\kappa_k(n^{-1}U_0^{*\top}U_0^{*})\leq \kappa_1(n^{-1}U_0^{*\top}U_0^{*})<\nu_2$. Furthermore, $C_d< {\sqrt{\nu_1}}$.
    \item[(C3)] The true parameters $(\alpha^{t*},\Delta^*)$ belong to $\mathcal{F}_1(U_0)$.
\end{itemize}
Condition (C0) assumes the dimension of the latent space is fixed, and
imposes an assumption on the order of the distance between $Z^{t*}$ and $U_0^*$, ensuring the closeness of the latent variables in the target network and transferable source networks. Condition (C1) concerns the closeness between $U_0$ and $U_0^{*}$ as well as the order of $\zeta_n$.
Condition (C2) assumes that the eigenvalues of $n^{-1}U_0^{*\top}U_0^*$ are bounded. This is mild and can also be found in \cite{zhang2020flexible, fan2024generalized, fan2025network}. Furthermore, $C_d<\sqrt{v_1}$ indicates that 
 $\|\Delta\|_*$ will less than $\sigma_{min}(U_{0}^{*})$ when $n$ is large. This assumption is justified because $\Delta$ represents a small deviation from $U_0$, and $U_0$ is close to $U_0^*$.

The following theorem provides an error bound for the final estimator $\hat Z^t$.

\begin{theorem}\label{thm1}
     Assume Conditions (C0)--(C3) hold.  
     If $\lambda\leq C_1n$ for some positive constant $C_1$, then there exists positive constants ${C}_0$, $\tilde{C}$, such that with 
a probability of at least $1-n^{-C_0}$,  \begin{equation*}
    \begin{aligned}
        \mathop{\text{min}}_{O: OO^\top=O^\top O=I_k} ||\hat{Z}^t-Z^{t*}O||_F\leq \tilde{C}.
    \end{aligned}
\end{equation*}
\end{theorem}

Theorem \ref{thm1} provides the upper bound of the final estimator $\hat Z^t$ and it implies that $n^{-1/2}$ $\mathop{\text{min}}_{O: OO^\top=O^\top O=I_k} ||\hat{Z}^t-Z^{t*}O||_F$ is dominated by $O_p(n^{-1/2})$. The orthogonal matrix arises in the theorem as a result of the fact that $Z^t$ is identifiable up to an orthogonal transformation. The detailed proof is given in the supplementary material.

Next, we examine the validity of the assumption in Condition (C1),  which concerns the closeness between $U_0^*$ and $\hat U_0$. For simplicity, we focus on the case when all source networks in transferable set share the same latent variables of the first $n$ nodes. 

Define a
feasible parameter space for source network datasets with known transferable set $\mathcal{A}$:
    \begin{equation*}
	\begin{aligned}
\mathcal{F}_2=\{&(U_0,\{\alpha^{s_l},U_l\}_{l=1}^{|\mathcal A|}): \mathring\Theta^{s_l}=\alpha^{s_l} 1_{N_l}^\top+1_{N_l}\alpha^{s_l\top}+(U_0^{\top},U_l^{\top})^\top(U_0^{\top},U_l^{\top}), \\
 &J_1U_0=U_0, J_2^lU_l=U_l,
 ||\mathring\Theta^{s_l}||_{\text{max}}<M_1,  \text{for}\ l\in\mathcal{A}\},
	\end{aligned}
\end{equation*}
for some finite positive constants $M_1$. And denote $(\hat U_0,\{\hat \alpha^{s_l},\hat U_l\}_{l=1}^{|\mathcal A|})$ as the minimizer of (\ref{2-2}) under parameter space $\mathcal{F}_2$. Consider the following condition:
\begin{itemize}
    \item [(C4)] The true parameters $(U_0^*,\{\alpha^{s_l*},U_l^*\}_{l=1}^{|\mathcal A|})$ belong to $\mathcal{F}_2$.
\end{itemize}
The following theorem provides an error bound for the estimator $\hat U_0$.

\begin{theorem}\label{thm2}
    Suppose all transferable source networks share the same $U_{0l}^*=U_0^{*}$, {and the sample sizes share the same order as $n$.} Assume that transferable network set $\mathcal{A}$ is known.  Under Condition (C2) and (C4), there exists positive constants $\tau_1$, $\tau_2$, and $\tilde{C}_1$, such that with a probability of at least $1-\sum_{l\in\mathcal{A}}N_l^{-\tau_1}-\sum_{l\in\mathcal{A}}n_l^{-\tau_1}-\exp(-\tau_2(2n+\lvert\mathcal{A}\rvert))$, 
    \begin{equation*}
       ||\hat{U_0}\hat{U_0}^\top-U_0^*U_0^{*\top}||_F\leq c_{\zeta}\sqrt{n},
    \mbox{ and }
         \mathop{\text{min}}_{O: OO^\top=O^\top O=I_k}||\hat{U_0}-U_0^*O||_*^2\leq \tilde{C}_1,
    \end{equation*}
    where $c_{\zeta}$ is defined in Theorem \ref{thm1}.
\end{theorem}
The proof of Theorem \ref{thm2} is provided in the supplementary material. The error bound of $\hat U_0$ is established, and it is shown that if the conditions in Theorem \ref{thm2} are satisfied and $\tilde{C}_1$ takes a value greater than $\xi^2$, then the estimator $\hat U_0$ satisfies the Condition (C1) in Theorem \ref{thm1}.
 \subsection{Transfer learning with detected transferable set}

In practice, the set of transferable networks is unknown.
Directly pooling all source networks for transfer learning may lead to worse performance compared to an estimation based solely on the target network, especially when some source networks provide little useful information.
This phenomenon is known as negative transfer and has been extensively studied in previous research \citep{torrey2010transfer,tian2023transfer}.
To mitigate the risk of negative transfer, it is essential to identify the set of transferable networks.

Inspired by \cite{tian2023transfer}, we propose a data-driven approach based on edge sampling \cite{li2020network}. In each sampling, 80\% of the edges from the target network are randomly selected for analysis. We apply Algorithm \ref{thm1} to transfer one source network to
analyze the adjusted target network
at a time. Based on the estimated probabilities, we predict the remaining 20\% of edges in the target network and compute the predictive loss.
The negative log-likelihood function is adopted as the loss function.
We identify a source network as transferable if its  predictive loss mean is not significantly greater than that of the baseline method.
More details are provided in Algorithm \ref{al4}. Algorithm \ref{al4} requires initial inputs of several hyperparameters.
{The value of $k$ is  predetermined and $\lambda^{[l](r)}$ can be selected using network cross-validation \citep{li2020network}.
The value of $\iota$ is a small positive constant, determined based on specific application scenarios.
Since the result is not sensitive to the selection of $R$, in all our following experiments, we take $R=3$. 
}
\begin{algorithm}[H]
	\caption{ Transferable set detection\label{al4}}
	\SetAlgoNoLine
	\SetKwInOut{Input}{\textbf{Input}}
	\SetKwInOut{Output}{\textbf{Output}}
	
	\Input{adjacency matrices of source networks $A^{s_l} \in \mathbb{R}^{N_l \times N_l}$ for $l=1,\ldots,L$; adjacency matrix of target network $A^t\in\mathbb{R}^{n\times n}$; latent space dimension $k$; a positive constant $\iota$;  $\lambda^{[l](r)}$ for $l=1,\ldots,L$ and $r=1,\ldots,R$.
	}
\Output{$\hat{\mathcal{A}}$.
	}
	\BlankLine
    \For {$r= 1, \cdots, R$}{
	Randomly sample 80\% of the node pairs $\Upsilon$ in the network $A^t$ \;
    Run Algorithm \ref{al3} on target network  on set $\Upsilon$ and $A^{s_l}$ with tuning parameter $\lambda^{[l](r)}$ for  all $ l=1,\ldots,L$, respectively, and calculate the loss function $\hat L_l^{(r)}$ on the
    held-out set $(A^t)_{ij}$ with $(i,j)\in\Upsilon^c$, where \begin{small}$\Upsilon^c=\{(i,j):(i,j)\notin \Upsilon\}$ \end{small}\;
   Fit vanilla latent space model on target network on set $\Upsilon$ and calculate the loss function $\hat L_0^{(r)}$ on $\Upsilon^c$\;
    }
	Calculate  \begin{footnotesize}$\hat L_0=1/R\sum_{r=1}^R\hat L_0^{(r)}$\end{footnotesize} and $\hat L_l=1/R\sum_{r=1}^R\hat L_l^{(r)}$ for $l=1,\ldots,L$, respectively. And compute  standard deviation as \begin{footnotesize}$\hat\sigma=\sqrt{\sum_{r=1}^R(\hat L_0^{(r)}-\hat L_0)^2/(R-1)}$\end{footnotesize}\;
\begin{footnotesize}

		\textbf{Output} $\hat{\mathcal{A}}=\{l\ne 0|\ \hat L_l-\hat L_0\leq \iota\hat\sigma\}$\end{footnotesize}.
\end{algorithm}

\section{Simulation studies}

In this section, we conduct multiple simulation studies to evaluate the performance of our proposed methods. 
We design the simulation settings as follows.
We set the latent space dimension to $k = 2$. 
The target network sizes are set to $n = 200,400$. The latent variables of the target network, denoted by $Z^{t*}$, are generated in three steps. For each $v = 1, \dots, k$, independently generate a vector $\mu_v \in \mathbb{R}^k$, where each element is drawn from the uniform distribution $\mathcal U[-1, 1]$.
Generate the matrix $\bar{Z}^t = (\bar{Z}^t_{ij}) \in \mathbb{R}^{n \times k}$ as follows. Randomly divide the $n$ nodes into $k$ subsets.  For each node $i = 1, \dots, n$, if it belongs to the $v$-th subset, then sample
   $
   \bar{Z}^t_{ij} \sim N(\mu_{vj}, 1), 
   $ for  $j = 1, \dots, k$.
Set $Z^{t*} = J_1 \bar{Z}^t$. The elements of 
$\alpha^{t*}$ are independently drawn from a uniform distribution $\mathcal U[-2.625, -0.875]$ to achieve a network density of approximately 8\%. The target network is generated by equation (\ref{s2-1}). 
The number of layers of the source networks is fixed at $L = 10$. The source network sizes are considered under three scenarios: (1) all source networks have the same size as the target network, i.e., $N_1 = \ldots = N_L = n$; (2) the number of nodes in the first five source networks is generated from  the uniform distribution $\mathcal U(n, 2n)$, while the remaining source networks have sizes fixed at $2n$; and (3) all source networks have $2n$ nodes.  The number of informative networks is set to $|\mathcal{A}| = 5$ or $10$. To generate the latent variables of the source networks, we first generate  diagonal matrices $D=\delta_{l}I_k$ for $l=1,\ldots, L$. Consider three cases for $\delta_l$: (i). $\delta_{l}=0$, for $l=1,\ldots,|\mathcal{A}|$,  and $ \delta_{l}=15$ for $l=|\mathcal{A}|+1,\ldots,L$; (ii).$\delta_{l}=5$, for $l=1,\ldots,|\mathcal{A}|$, and $ \delta_{l}=15$ for $l=|\mathcal{A}|+1,\ldots,L$;
(iii). $\delta_{l} \sim U(0,5)$ for $l=1,\ldots,|\mathcal{A}|$ and $\delta_{l} \sim U(10,15)$ for $l=|\mathcal{A}|+1,\ldots,L$.
We then generate the matrix $V_1 \in \mathbb{R}^{n \times k}$ as the first $k$ columns of a random orthogonal matrix and the matrix $V_2 \in \mathbb{R}^{k \times k}$ as another random orthogonal matrix, both generated using the ``randortho" function in the R package "pracma" \citep{borchers2019package}. The final form of ${U}_{0l}^*$ is given by ${U}_{0l}^* = J_1 V_1DV_2^\top + Z^{t*}$, which yields a value of $\delta^*$ equal to $k\delta_l$.
If $N_l > n$, an additional matrix $U_l^* \in \mathbb{R}^{(N_l - n) \times k}$ is generated in a similar process as that of $Z^{t*}$: the $N_l - n$ nodes are divided into $k$ subsets, and for each $i$-th node in subset $v$, $\bar U_{l,ij} \sim N(\mu_{vj}, 1)$ is generated for $i = 1, \ldots, N_l - n$. Finally, $U_l^* = J_2^l \bar U_l$. The latent variable for the $l$-th layer of the source networks is then given by $Z^{s_l*} = ({U}_{0l}^{*\top}, U_l^{*\top})^\top$.
For the degree heterogeneity parameters,
$\alpha^{s_l*}$ for $l = 1, \dots, L$ is elemently drawn from a uniform distribution $\mathcal U[-1.313, -0.438]$ to maintain a network density of approximately 21\%.
The source networks are generated by equation (\ref{equ S}).

We compare the proposed method with four alternative approaches. (1) TLK: which applies transfer learning with the set of informative networks known.
(2) TLB: which applies transfer learning utilizing all source networks regardless of their informativeness  (with "B" indicating blink); (3) TLE: which transfers knowledge from subnetworks of the source networks that only contain the nodes present in the target network, and the procedure of transferable detection is also applied  (with "E" indicating equal); and (4) one-mode, which estimates the target network without incorporating information from source networks.
Several metrics are reported to evaluate the performance of the proposed methods. These include the relative error of  $Z^t$, denoted as $\Delta_Z=\left\|\hat{Z}^t \hat{Z}^{t\top}-Z^{t*} Z^{t* \top}\right\|_F^2 /\left\|Z^{t*}Z^{t* \top}\right\|_F^2$, the relative error of  $\alpha^t$, denoted as $\Delta_{\alpha}=\left\|\hat{\alpha}^t-\alpha^{t*}\right\|_2^2 /\left\|\alpha^{t*}\right\|_2^2$, and the relative error of $\Theta^t$, denoted as $\Delta_{\Theta}=\left\|\hat{\Theta}^t-\Theta^{t*}\right\|_F^2 /\left\|\Theta^{t*}\right\|_F^2$. 
Additionally, we compute the true positive rate (TPR) and false positive rate (FPR) for the detection of informative source networks.

All experiments are run on a computer equipped with Intel(R) Xeon(R) E5-2643 v4 CPUs and 256GB of RAM. 
The simulation results are provided in Tables  \ref{tab2}, \ref{tab3} below and the remaining results are in the supplementary material. The performance of 
TLD method is better than the TLB, TLE and one-mode methods and  comparable to TLK method, when noninformative networks exist.
For example, the values of $\Delta_Z$
are 0.0436, 0.0436, 0.0529, 0.1383, 0.1772 for the TLK, TLD, TLE, TLB, and one-mode, respectively,
when $n=200$, $|\mathcal A|=5$ of case (i) in scenario (2).When all networks are informative, TLK, TLD, and TLB methods perform similarly, showing significant improvement compared to the TLE and one-mode estimations.
Moreover, as the value of  $\delta$ of informative networks decreases,  the performance of all transfer learning methods improves. 
This result is expected, as the more similar the latent variables of the transferable source network and the target network are, the more informative the source network becomes for estimation.
An increase in the number of informative networks, $|\mathcal{A}|$, also enhances the transfer performance. 
Furthermore, as the sizes of the source networks increase across the three proposed scenarios, the estimation errors decrease.
As the size of the target network grows, the performance of the proposed methods improves.
This is because having more nodes enhances the amount of information available. In addition,
TLD identifies the transferable networks,  thereby mitigating the influence of noninformative networks and preventing negative transfer.  This is supported by Table \ref{tab3}, which shows that the TPR in all settings is nearly 1, while the FPR remains close to 0 in settings where $|\mathcal{A}|=5$. 


\section{Real data analysis}
We apply the proposed method to two real-world examples. One dataset consists of trade networks among countries and regions. The other dataset is the POLECAT political event networks, which records geopolitical interactions among global entities.
To assess the performance of the proposed method, we perform random sampling over network edges. Specifically,
We randomly remove a proportion of the edges in the target network,
and train the model using the remaining edges to predict the missing linkages. The missing ratio is $p$. The Brier score is calculated as the performance measure, and the experiments are repeated 50 times.

\subsection{FAO trade networks}

\begin{figure}[H]%
     \begin{center}
    \includegraphics[width=0.7\textwidth]{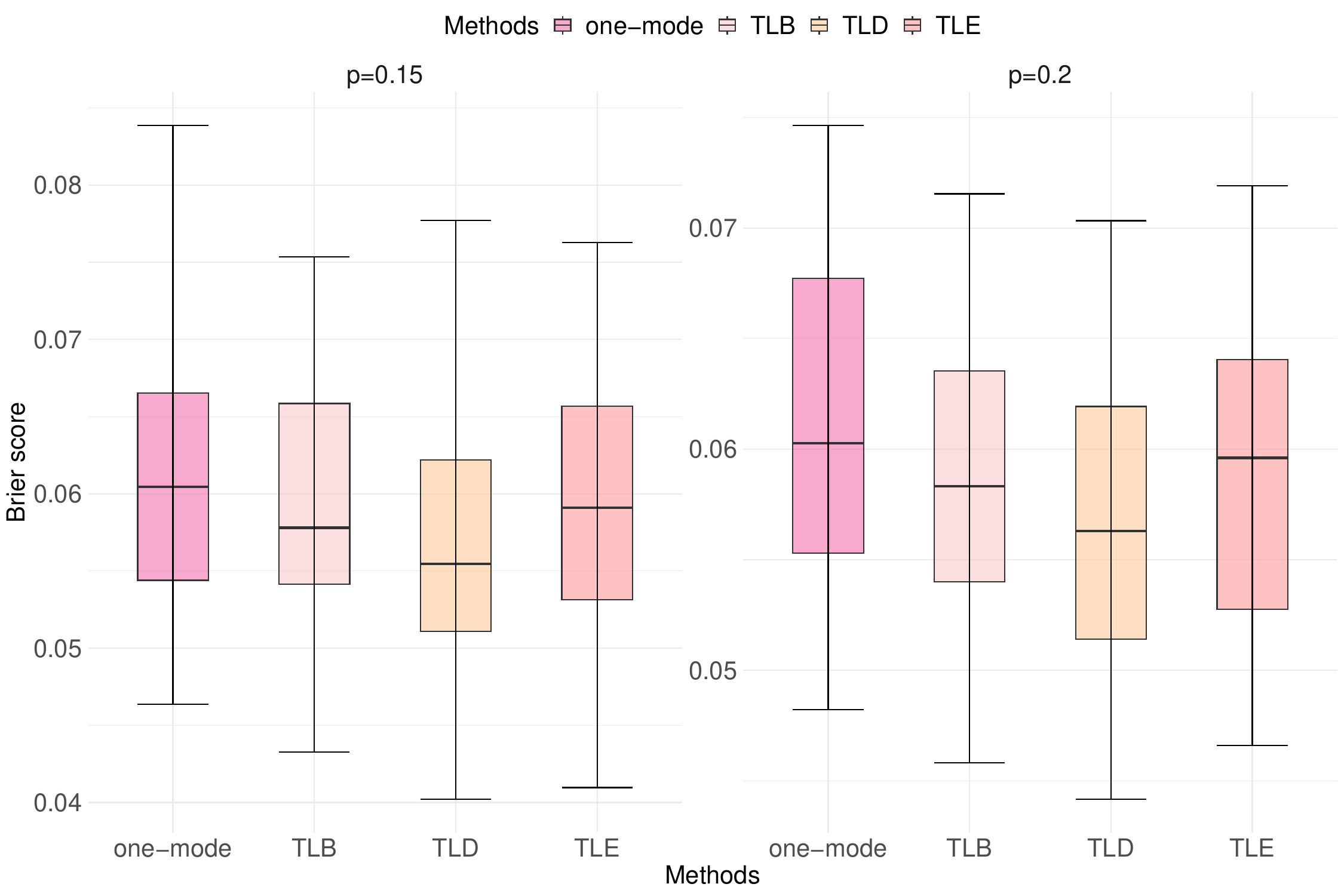}
  \end{center}
\caption{The prediction results of FAO trade network}
    \label{fig1}
\end{figure}

FAO trade network dataset, compiled by the Food and Agriculture Organization of the 
United Nations and organized by \cite{de2015structural}, captures global food trade flows and is publicly available at http://www.fao.org. It includes the import and export relationships between 214 countries and regions in 2010, with 318,346 connections recorded.
\begin{table}[H]
    \centering
	\renewcommand{\arraystretch}{0.8}
	\caption{The means and standard deviations of the relative estimation errors under scenario (2). (i), (ii) and (iii) correspond to the 3 cases of $\delta$ generation. The performance of one-mode remains consistent across all cases, is therefore shown only once. In each cell:  mean (sd).\\}
	\label{tab2}
	\resizebox{!}{0.4\textwidth}{
		\begin{tabular}{cccccccc|cccccc}
			\hline\hline
            ~ & ~ &\multicolumn{6}{c|}{$|\mathcal{A}|=5$}&\multicolumn{6}{c}{$|\mathcal{A}|=10$}\\
        ~ & ~ & ~ &  $n=200$ & ~ & ~ & $n=400$ & ~ & ~ & $ n=200 $& ~ & ~ & $n=400$&~ \\ \hline
        ~ & ~ & $\Delta_Z$ & $\Delta_\Theta$ & $\Delta_\alpha$ & $\Delta_Z$ & $\Delta_\Theta$ & $\Delta_\alpha$ & $\Delta_Z$ & $\Delta_\Theta$ & $\Delta_\alpha$ & $\Delta_Z$ & $\Delta_\Theta$ & $\Delta_\alpha$  \\
        (i) & TLK & 0.0436  & 0.0337  & 0.0467  & 0.0327  & 0.0316  & 0.0441  & 0.0317  & 0.0305  & 0.0450  & 0.0278  & 0.0302  & 0.0433  \\ 
        ~ & ~ & (0.0177) & (0.0050) & (0.0040) & (0.0089) & (0.0030) & (0.0024) & (0.0106) & (0.0032) & (0.0029) & (0.0064) & (0.0026) & (0.0024) \\ 
        ~ & TLD & 0.0436  & 0.0337  & 0.0467  & 0.0327  & 0.0316  & 0.0441  & 0.0317  & 0.0305  & 0.0450  & 0.0278  & 0.0302  & 0.0433  \\ 
        ~ & ~ & (0.0177) & (0.0050) & (0.0040) & (0.0089) & (0.0030) & (0.0024) & (0.0106) & (0.0032) & (0.0029) & (0.0064) & (0.0026) & (0.0024) \\ 
        ~ & TLE & 0.0529  & 0.0359  & 0.0478  & 0.0391  & 0.0333  & 0.0450  & 0.0461  & 0.0345  & 0.0473  & 0.0351  & 0.0323  & 0.0446  \\ 
        ~ & ~ & (0.0153) & (0.0048) & (0.0038) & (0.0083) & (0.0031) & (0.0026) & (0.0132) & (0.0040) & (0.0033) & (0.0072) & (0.0029) & (0.0026) \\ 
        ~ & TLB & 0.1383  & 0.0591  & 0.0597  & 0.0853  & 0.0466  & 0.0526  & 0.0317  & 0.0305  & 0.0450  & 0.0278  & 0.0302  & 0.0433  \\ 
        ~ & ~ & (0.0076) & (0.0030) & (0.0029) & (0.0059) & (0.0016) & (0.0015) & (0.0106) & (0.0032) & (0.0029) & (0.0064) & (0.0026) & (0.0024) \\ 
        ~ & one-mode & 0.1772  & 0.0591  & 0.0524  & 0.1162  & 0.0503  & 0.0511  & 0.1772  & 0.0591  & 0.0524  & 0.1162  & 0.0503  & 0.0511  \\ 
        ~ & ~ & (0.0098) & (0.0018) & (0.0022) & (0.0038) & (0.0009) & (0.0013) & (0.0098) & (0.0018) & (0.0022) & (0.0038) & (0.0009) & (0.0013) \\ 
        (ii) & TLK & 0.0774  & 0.0417  & 0.0498  & 0.0519  & 0.0366  & 0.0465  & 0.0601  & 0.0381  & 0.0489  & 0.0438  & 0.0347  & 0.0459  \\ 
        ~ & ~ & (0.0111) & (0.0038) & (0.0037) & (0.0104) & (0.0041) & (0.0034) & (0.0121) & (0.0039) & (0.0033) & (0.0071) & (0.0029) & (0.0026) \\ 
        ~ & TLD & 0.0774  & 0.0417  & 0.0498  & 0.0519  & 0.0366  & 0.0465  & 0.0601  & 0.0381  & 0.0489  & 0.0438  & 0.0347  & 0.0459  \\ 
        ~ & ~ & (0.0111) & (0.0038) & (0.0037) & (0.0104) & (0.0041) & (0.0034) & (0.0121) & (0.0039) & (0.0033) & (0.0071) & (0.0029) & (0.0026) \\ 
        ~ & TLE & 0.0898  & 0.0448  & 0.0516  & 0.0569  & 0.0376  & 0.0468  & 0.0657  & 0.0389  & 0.0490  & 0.0506  & 0.0364  & 0.0467  \\ 
        ~ & ~ & (0.0100) & (0.0031) & (0.0031) & (0.0061) & (0.0023) & (0.0020) & (0.0066) & (0.0028) & (0.0029) & (0.0067) & (0.0025) & (0.0021) \\ 
        ~ & TLB & 0.1499  & 0.0616  & 0.0605  & 0.0933  & 0.0487  & 0.0536  & 0.0601  & 0.0381  & 0.0489  & 0.0438  & 0.0347  & 0.0459  \\ 
        ~ & ~ & (0.0123) & (0.0034) & (0.0027) & (0.0037) & (0.0016) & (0.0019) & (0.0121) & (0.0039) & (0.0033) & (0.0071) & (0.0029) & (0.0026) \\ 
        (iii) & TLK & 0.0575  & 0.0366  & 0.0472  & 0.0388  & 0.0328  & 0.0443  & 0.0389  & 0.0322  & 0.0458  & 0.0339  & 0.0320  & 0.0443  \\ 
        ~ & ~ & (0.0118) & (0.0037) & (0.0033) & (0.0082) & (0.0029) & (0.0023) & (0.0162) & (0.0041) & (0.0028) & (0.0095) & (0.0032) & (0.0025) \\ 
        ~& TLD & 0.0580  & 0.0367  & 0.0472  & 0.0431  & 0.0339  & 0.0449  & 0.0389  & 0.0322  & 0.0458  & 0.0339  & 0.0320  & 0.0443  \\ 
        ~ & ~ & (0.0134) & (0.0041) & (0.0033) & (0.0084) & (0.0029) & (0.0024) & (0.0162) & (0.0041) & (0.0028) & (0.0095) & (0.0032) & (0.0025) \\ 
        ~ & TLE & 0.0710  & 0.0400  & 0.0491  & 0.0505  & 0.0363  & 0.0467  & 0.0485  & 0.0348  & 0.0473  & 0.0389  & 0.0335  & 0.0454  \\ 
        ~ & ~ & (0.0118) & (0.0034) & (0.0027) & (0.0099) & (0.0030) & (0.0022) & (0.0087) & (0.0030) & (0.0029) & (0.0069) & (0.0028) & (0.0024) \\ 
        ~ & TLB & 0.1269  & 0.0555  & 0.0571  & 0.0820  & 0.0459  & 0.0525  & 0.0389  & 0.0322  & 0.0458  & 0.0339  & 0.0320  & 0.0443  \\ 
        ~ & ~ & (0.0060) & (0.0023) & (0.0028) & (0.0093) & (0.0030) & (0.0023) & (0.0162) & (0.0041) & (0.0028) & (0.0095) & (0.0032) & (0.0025) \\ 
		\hline
			
		\hline
	\end{tabular}}
\end{table}We construct undirected networks by establishing an edge between two countries and regions if a trade relationship exists between them, following the preprocessing method of \cite{jing2021community}. 
The beet pulp network is used as the target due to its small sample size and low density (75 nodes, density 0.0789).
Six auxiliary networks are selected—Food preparations not elsewhere specified, Beer of barley, Crude materials, Bread, Chicken meat, and Pastry—all of which cover the nodes of the beet pulp network. These auxiliary networks have sample sizes ranging from 148 to 202 and densities ranging from 0.0959 to 0.2378.
The latent space dimension is fixed at 2  for visualization.
We compared the proposed TLD method and TLE,TLB, and one-mode methods using random sampling over network edges. 
As shown in Figure \ref{fig1}, TLD consistently outperforms others under both missing ratios, benefiting from transferable auxiliary information. The gap between TLD and TLB highlights the importance of identifying transferable networks.
Finally, we fit TLD on the full data, selecting Beer of barley, Bread, and Chicken meat as transferable due to shared supply patterns. Clustering results based on $\hat Z^t$ yield four groups, visualized in supplementary material. These groupings align with the characteristics of the countries and regions and offer insights for formulating trade strategies.

\subsection{POLECAT political event network}
We analyze the POLECAT political event network dataset \cite{halterman2023plover, baum2024doubly}, which records 670,322 geopolitical interactions among 199 entities from Jan 2023 to Apr 2024, across 16 event types (e.g., concede, protest, aid). Sixteen undirected networks are constructed accordingly.

\begin{figure}[H]%
      \begin{center}
    \includegraphics[width=0.65\textwidth]{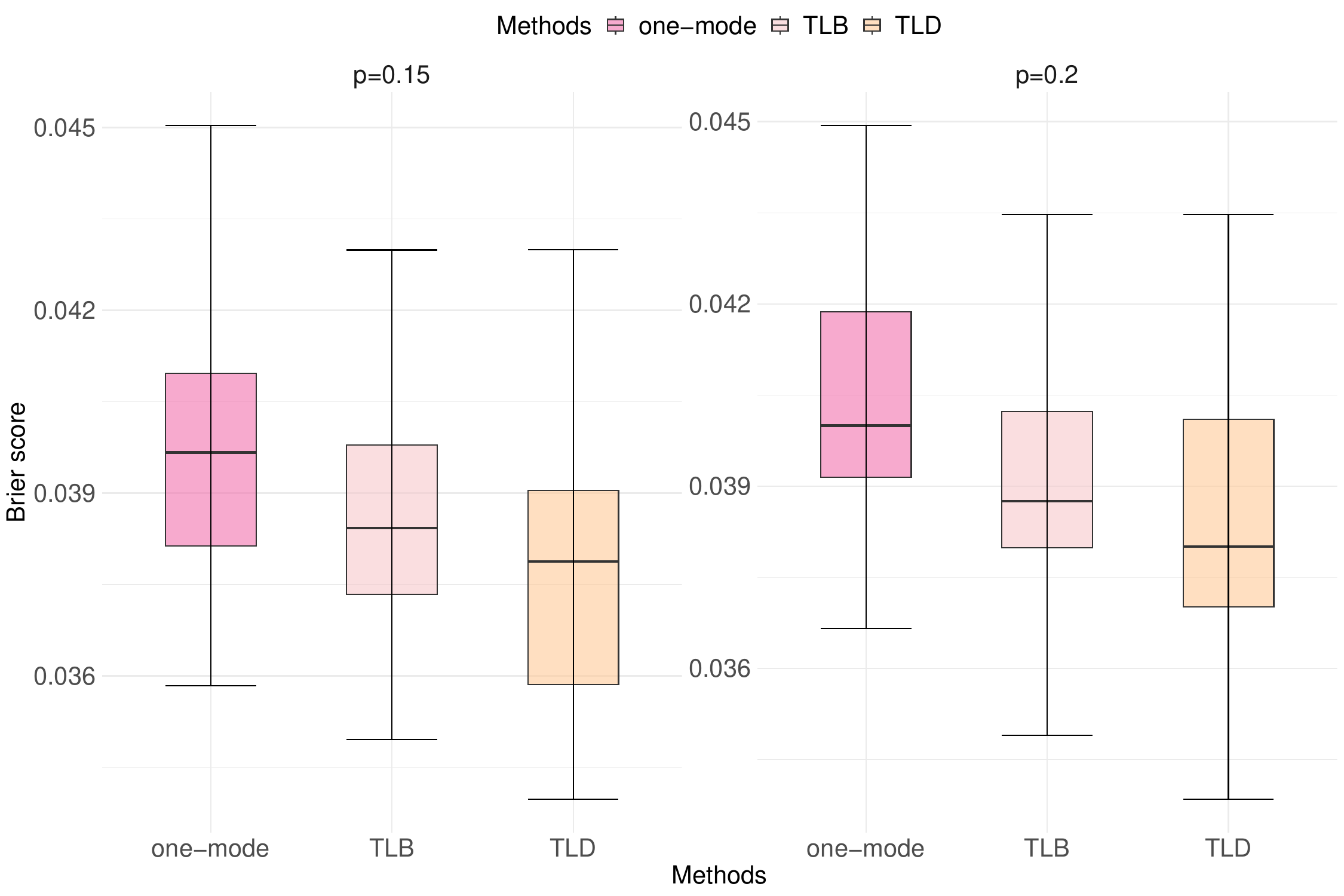}
  \end{center}
\caption{The prediction results of POLECAT network}
    \label{fig2}
\end{figure}

We take the lowest-density network, concede (density 0.0641), as the target and treat the remaining networks as source networks.
The latent space dimension is fixed at 2.
 The results of Brier score based on random sampling
 for TLD and three compared methods is provided in Figure \ref{fig2}.
   TLD achieves improved Brier scores over the others, validating effective knowledge transfer and the benefit of identifying transferable networks.
Finally, we fit TLD on the full data. For the network of concessions, the transferable networks include those of protests, assaults, and threats.
These political events are either causes or consequences of concessions \citep{langner2001motivational,chen2021statistical}.

Other networks such as those of assaults, sanctions, and protests are also taken as target network, with the corresponding remaining 15 networks serving as auxiliary networks for each target. The transferable networks selected for each target network are presented in the supplementary material. 

\section{Conclusion}
Precise estimation of latent variables is crucial for downstream tasks such as link prediction and community detection under latent space model. In this work, we propose a transfer learning method that uses information from source networks to improve the latent variable estimation in the target network.
We design a two-stage algorithm and investigate its identification conditions and estimation procedure at each stage. Additionally, we derive the error bound for the final estimator of latent variables in the target network. To mitigate negative transfer, we develop an algorithm to detect transferable sets of source networks.
Finally, we demonstrate the effectiveness of our method through extensive simulations and two real-world data analyses.

There are several directions worth investigating for future research. First, it is theoretically worthwhile to  explore how source networks reduce the upper bound of latent variable estimation error.
Second, the proposed framework could be extended to a distributed transfer learning setting, where source networks are stored across different computing nodes, and privacy protection becomes a relevant concern.

\begin{table}[H]
    \centering
	\renewcommand{\arraystretch}{0.9}
	\caption{The means and standard deviations of TPR and FPR of the transferable network detection. (i), (ii) and (iii) correspond to the three different cases of the generation of $\delta$, and (1), (2) and (3) indicates the three source networks size scenarios. In each cell:  mean (sd).\\}
	\label{tab3}
	\resizebox{!}{0.23\textwidth}{
		\begin{tabular}{cccccccc|cccccc}
			\hline\hline
            ~ & $|\mathcal{A}|$ &\multicolumn{6}{c|}{$n=200$} &\multicolumn{6}{c}{$n=400$} \\
            ~ & ~ & ~ & TPR & ~ & ~ & FPR & ~ & ~ & TPR & ~ & ~ & FPR & ~ \\
             \cline{3-14}
            ~ & ~ & (i) & (ii) & (iii) & (i) & (ii) & (iii) & (i) & (ii) & (iii) & (i) & (ii) & (iii) \\
            (1) & 5 & 1.0000 & 1.0000 & 1.0000 & 0.0000 & 0.0000 & 0.0810 & 1.0000 & 1.0000 & 1.0000 & 0.0000 & 0.0000 & 0.0750 \\ 
            ~ & ~ & (0.0000) & (0.0000) & (0.0000) & (0.0000) & (0.0000) & (0.0954) & (0.0000) & (0.0000) & (0.0000) & (0.0000) & (0.0000) & (0.0851) \\ 
            ~ & 10 & 1.0000 & 1.0000 & 1.0000 & (-) & (-) & (-) & 1.0000 & 1.0000 & 1.0000 & (-) & (-) & (-) \\ 
            ~ & ~ & (0.0000) & (0.0000) & (0.0000) & (-) & (-) & (-) & (0.0000) & (0.0000) & (0.0000) & (-) & (-) & (-) \\ 
            (2) & 5 & 1.0000 & 1.0000 & 1.0000 & 0.0000 & 0.0000 & 0.0310 & 1.0000 & 1.0000 & 1.0000 & 0.0000 & 0.0000 & 0.0833 \\ 
            ~ & ~ & (0.0000) & (0.0000) & (0.0000) & (0.0000) & (0.0000) & (0.0788) & (0.0000) & (0.0000) & (0.0000) & (0.0000) & (0.0000) & (0.0855) \\ 
            ~ & 10 & 1.0000 & 1.0000 & 1.0000 & (-) & (-) & (-) & 1.0000 & 1.0000 & 1.0000 & (-) & (-) & (-) \\ 
            ~ & ~ & (0.0000) & (0.0000) & (0.0000) & (-) & (-) & (-) & (0.0000) & (0.0000) & (0.0000) & (-) & (-) & (-) \\ 
            (3) & 5 & 1.0000 & 1.0000 & 1.0000 & 0.0000 & 0.0000 & 0.0500 & 1.0000 & 1.0000 & 1.0000 & 0.0000 & 0.0000 & 0.0917 \\
            ~ & ~ & (0.0000) & (0.0000) & (0.0000) & (0.0000) & (0.0000) & (0.0784) & (0.0000) & (0.0000) & (0.0000) & (0.0000) & (0.0000) & (0.0851) \\ 
            ~ & 10 & 1.0000 & 1.0000 & 1.0000 & (-) & (-) & (-) & 1.0000 & 1.0000 & 1.0000 & (-) & (-) & (-) \\ 
            ~ & ~ & (0.0000) & (0.0000) & (0.0000) & (-) & (-) & (-) & (0.0000) & (0.0000) & (0.0000) & (-) & (-) & (-) \\ 
    	\hline
			
		\hline
	\end{tabular}}
\end{table}

\bibliography{Dissertation_full}

\end{document}